\documentstyle[prl,aps]{revtex}

\begin{document}
\draft
\title{Exact Results for a Kondo Problem in One Dimensional t-J Model}
\author{Yupeng Wang$^a$ \and Jianhui Dai$^b$ \and Zhanning Hu$^d$ 
\and Fu-Cho Pu$^{c,d}$}
\address{$^a$Cryogenic Laboratory, Chinese Academy of Sciences, 
Beijing 100080, P. R. China\\
$^b$Zhejiang Institute of Modern Physics, Zhejiang University, 
Hangzhou 310027, P. R. China\\
$^c$Department of Physics, Guangzhou Teacher College, 
Guangzhou 510400, P. R. China\\
$^d$Institute of Physics, Chinese Academy of Sciences, 
Beijing 100080, P. R. China}
\maketitle
\begin{abstract}
We propose an integrable Kondo problem in a one-dimensional (1D) $t-J$ 
model. With the open boundary condition of the wave functions at the 
impurity sites, the model can be exactly solved via Bethe ansatz for a 
class of $J_{R,L}$ (Kondo coupling constants) and $V_{L,R}$ (impurity
potentials) parametrized by a single parameter $c$. The integrable value 
of $J_{L,R}$ runs from negative infinity to positive infinity, which 
allows us to study both the ferromagnetic Kondo problem and 
antiferromagnetic Kondo problem in a strongly correlated electron 
system. Generally, there is a residual entropy for the ground state, 
which indicates a typical non-Fermi liquid behavior. 
\end{abstract}
Recently, considerable attention has been drawn by the theory of 
impurities in both the Fermi and Luttinger liquids, and many new 
developments have been addressed. This renewed interest in the 
quantum impurity problems was partially stimulated by the search for
non-Fermi-liquid fixed points beyond the well known Luttinger liquid
 and its plausible  relevance to the theory of heavy fermions and 
cuprate superconductors. A great progress is the study on the 
multi-channel Kondo problem, which has been extensively studied by 
the conformal field theory in the presence of boundary\cite{1,2} and 
Bethe ansatz\cite{3,4,5}. These powerful methods allow to obtain the 
low temperature thermodynamics near the critical point. Using 
simpler bosonization and renormalization group techniques, Kane and 
Fisher have shown that a potential scatter center embedded in a Luttinger
 liquid is driven to a strong-coupling fixed point by the repulsive
 electron-electron interactions\cite{6}. This is the first time to show 
that a single impurity in a Luttinger liquid behaves rather different from
 that in a Fermi liquid, and directly stimulates the study on the problem
 of local perturbations to a Luttinger liquid and especially on the Kondo
 problem in a Luttinger liquid. The Luttinger-Kondo problem was first
 considered by Lee and Toner\cite{7}, who found the crossover of the
 Kondo temperature from power law dependence on the Kondo coupling 
constant to an exponential one. Subsequently, a poor man's scaling
 was performed by Furusaki and Nagaosa\cite{8}, who addressed a 
conjecture which states that ferromagnetic Kondo screening may occur
 in a Luttinger liquid. The boundary conformal field theory\cite{9} gave 
out a classification of critical behavior for a Luttinger liquid coupled 
to a magnetic impurity (without impurity potential). It turns out that 
there are only two possibilities, a local Fermi liquid with standard 
low-temperature thermodynamics or a non-Fermi liquid observed by 
Furusaki and Nagaosa. It is now clear that the non-Fermi-liquid behavior 
is induced by the tunneling effect of conduction electrons through the
 impurity which depends only on the bulk properties but not on the 
detail of the impurity\cite{10,11}.
\par
Despite these important progress, the problem of few impurities (potential,
 magnetic, especially both) embedded in a strongly correlated system is 
still not well understood. We remark that there are a few progress
 related to this problem: A spin $S>1/2$ impurity in a spin $1/2$
 Heisenberg chain solved many years ago by Andrei and 
Johannesson\cite{12} and generalized to arbitrary spins by Lee
 and Schlottmann, and Schlottmann\cite{13}; an integrable impurity 
in the supersymmetric $t-J$ model\cite{14} with a very complicated
 hamiltonian solved by Bed{\"u}rftig et al. We remark here that 
these models are somewhat special for the absence of backward 
scattering off the impurity, and therefore are hard to give a 
deep understanding for the Kondo problem in a Luttinger liquid, where 
the backward scattering is crucial to the fixed point of the system. 
In addition, Wang and Voit has proposed an integrable model of a 
single magnetic impurity in a $\delta-$potential Fermi gas\cite{11} with a 
special value of Kondo coupling constant.
\par
Attempting to understand effects of magnetic impurities in a strongly 
correlated electron system, we study the properties of the integrable $t-J$

model coupled to two magnetic impurities sited at the ends of the system. 
Our starting point is the following hamiltonian

\begin{eqnarray}
H=H_0+H_i,\nonumber\\
H_0=-\sum_{j=1,\sigma}^{N_a-1}(C_{j\sigma}^\dagger
 C_{j+1\sigma}+h.c.)+2\sum_{j=1}^{N_a-1}[{\bf S}_j\cdot 
{\bf S}_{j+1}+Vn_jn_{j+1}],\\
H_i=J_L{\bf S}_1\cdot {\bf S}_L+V_Ln_1+J_R{\bf S}_{N_a}\cdot 
{\bf S}_R+V_Rn_{N_a},\nonumber
\end{eqnarray}
where $C_{j\sigma} (C_{j\sigma}^\dagger)$ are annihilation (creation)
 operators of the conduction electrons; $V$, $J_{R,L}$, $V_{R,L}$ are 
the nearest neighbor interaction constant, the Kondo coupling constants 
and the impurity potentials respectively;
 ${\bf S}_j=\frac12\sum_{\sigma,\sigma'}C_{j\sigma}^\dagger
{\bf \sigma}_{\sigma\sigma'}C_{j\sigma'}$ is the spin
 operator of the conduction electrons; ${\bf S}_{L,R}$ are the
 local moments with spin-$1/2$ sited at the left and right end of
 the system respectively; $n_j=C_{j\uparrow}^\dagger 
C_{j\uparrow}+C_{j\downarrow}^\dagger C_{j\downarrow}$ are the number
 operator of conduction electrons; $N_a$ is the length or site number
 of the system. Notice that the single occupation condition $n_j\leq 1$ 
is assumed for the hamiltonian (1). We remark that the model is very
 resonable for the absence of redundant terms in the hamiltonian.

\par
It is well known that $H_0$ is exactly solvable for $V=-1/4,
3/4$\cite{15,16}. In this letter, we only study $V=3/4$ case while both
 the charge- and spin-sector can be described by a Luttinger liquid. The
 $V=-1/4$ (supersymmetric $t-J$ model) can be followed without any 
difficulty. By including the impurities, any electron impinging on the
 boundaries will be completely reflected and suffer a reflection matrix 
$R_{j,L}$ or $R_{j,R}$\cite{11}. The waves are therefore reflected at
 either end as
\begin{eqnarray}
e^{ik_jx}\to R_{j,L}(k_j)e^{-ik_jx},\;\;\;  x\sim 1,\nonumber\\
e^{ik_jx}\to R_{j,R}^{-1}(k_j)e^{-ik_jx-2ik_jN_a},\;\;\; x\sim N_a
\end{eqnarray}
The reflecting Yang-Baxter equation\cite{17,18}

\begin{eqnarray}
S_{jl}(q_j-q_l)R_{j,a}(q_j)S_{jl}(q_j+q_l)R_{l,a}(q_l)=R_{l,a}(q_l)
S_{jl}(q_j+q_l)R_{j,a}(q_j)S_{jl}(q_j-q_l),\;\; a=R,L,
\end{eqnarray}
constrains the integrablity of the present model. Here $S_{jl}$ is the
 electron-electron scattering matrix in the bulk. By evaluating eq.(3),
 we obtain that the present model is exactly solvable with the following
 parametrized $J_a$ and $V_a$ ($a=L,R$): $J_a=-8/(2c_a+3)(2c_a-1)$,
 $V_a=(4c_a^2-7)/(2c_a+3)(2c_a-1)$, where $c_{L,R}$ are two arbitrary
 real constants. Generally, $J_L$ and $J_R$ may take different values. 
Here we consider only $J_R=J_L$, $V_R=V_L$, i.e., $c_L=c_R=c$ case. 
The solution of our model in the integrable line is quite similar
 to those of other integrable models with open boundaries\cite{17,18}. 
Notice that the reflection matrix $R_{L,R}$ ($K_\pm$ in ref.[17]) is an
 operator one rather than a c-number matrix. The spectra of the 
hamiltonian (1) are uniquely determined by the following Bethe ansatz
 equations
\begin{eqnarray}
(\frac{q_j-\frac i2}{q_j+\frac
i2})^{2N_a}=[\frac{q_j+i(c-1)}{q_j-i(c-1)}]^2
\prod_{r=\pm1}\{\prod_{l\neq j}^N\frac{q_j-rq_l-i}{q_j-rq_l+i}
\prod_{\alpha=1}^M\frac{q_j-r\lambda_\alpha+
\frac i2}{q_j-r\lambda_\alpha-\frac i2}\}\nonumber\\
\{\frac{\lambda_\alpha-i(c-\frac12)}{\lambda_\alpha+
i(c-\frac12)}\frac{\lambda_\alpha+i(c+\frac12)}
{\lambda_\alpha-i(c+\frac12)}\}^2
\prod_{r=\pm1}\prod_{j=1}^N\frac{\lambda_\alpha-rq_j+
\frac i2}{\lambda_\alpha-rq_j-\frac
i2}=\prod_{r=\pm1}\prod_{\beta\neq\alpha}^M
\frac{\lambda_\alpha-r\lambda_\beta+i}
{\lambda_\alpha-r\lambda_\beta-i},
\end{eqnarray}
with the eigenvalue of (1) given by $E=2N-\sum_{j=1}^N4/(4q^2_j+1)$. Here 
$N$ is the number of conduction electrons, $q_j=1/2\tan(k_j/2)$, and $k_j$
 and $\lambda_\alpha$ are the rapidities of charge and spin respectively. 
$M$ is the number of spin-down electrons.
\par
Below we discuss the ground state properties for different regions of 
parameter $c$.
\par
(i) $c\geq 1$. The system falls into the ferromagnetic Kondo coupling
 regime and no bound state can exist at low energy scales. The ground
 state is thus described by two sets of real parameters $\{q_j\}$ and
 $\{\lambda_\alpha\}$. Define the quantities
\begin{eqnarray}
Z_{N_a}^c(q)=\frac1\pi\{-\theta_1(q)+
\frac 1{2N_a}[\phi_c(q)-\sum_{\alpha=-M}^M
\theta_1(q-\lambda_\alpha)+\sum_{j=-N}^N\theta_2(q-q_j)]\},\nonumber\\
Z_{N_a}^s(\lambda)=\frac
1{2N_a}\{\phi_s(\lambda)-\sum_{j=-N}^N\theta_1(\lambda-q_j)+
\sum_{\alpha=-M}^M\theta_2(\lambda-\lambda_\alpha)\},
\end{eqnarray}
with the phase shifts $\phi_c(q)=-\theta_2(q)+4\tan^{-1}[q/(c-1)]$,
$\phi_s(\lambda)=-\theta_2(\lambda)+4\tan^{-1}
[\lambda/(c+1/2)]-4\tan^{-1}[\lambda/(c-1/2)]$  
induced by the impurity in the charge- and spin-sector respectively,
 and $\theta_n(x)=-2\tan^{-1}(2x/n)$. Note above we have used the 
reflection symmetry of the Bethe ansatz equations to include solutions
 with $q_{-j}=-q_j$ and $\lambda_{-\alpha}=-\lambda_\alpha$.
 The Bethe ansatz equations are solved by $Z_{N_a}^c(q_j)=I_j/{N_a}$ 
and $Z_{N_a}^s(\lambda)=J_\alpha/{N_a}$, where $I_j$ and $J_\alpha$ 
are non-zero integers. In the ground state, $I_j$ and $J_\alpha$ must
 be consecutive numbers around zero symmetrically to minimize the 
energy. The roots $q_j$ and $\lambda_\alpha$ becomes dense in the
 thermodynamic limit, and we define their densities as

\begin{eqnarray}
\rho_{N_a}^c(q)=\frac{dZ_{N_a}^c(q)}{dq},{~~~~~}\rho_{N_a}^s(\lambda)
=\frac{dZ_{N_a}^s(\lambda)}{d\lambda}.
\end{eqnarray}
The cutoffs of $q$ and $\lambda$ in the ground state are $\pm Q$ and 
$\pm \Lambda$ respectively, which correspond to 
$Z_{N_a}^c(\pm Q)=\pm(N+1/2)/{N_a}$,
 $Z_{N_a}^s(\pm\Lambda)=\pm(M+1/2)/{N_a}$. In the thermodynamic 
limit $N_a\to \infty$, $N\to\infty$, and $N/{N_a}\to finite$,
 we find that the energy is minimized at $\Lambda\to \infty$, 
which gives a result of $N=2M$ by integrating eq.(6). This 
leaves a spin triplet ground state which is contradicting to 
the Furusaki-Nagaosa conjecture\cite{8}. Notice that the 
magnetization is given by $1/2(N+2-2M)$.
\par
(ii) $1/2<c<1$. The system falls also into the ferromagnetic Kondo 
coupling regime. However, unlike case (i), two bound states of electrons
 can be formed around the impurities in the ground state, which 
correspond to two imaginary $q$ modes at $q=\pm i(1-c)$. The
 Bethe ansatz equations for the real modes are thus reduced to

\begin{eqnarray}
(\frac{q_j-\frac i2}{q_j+\frac
i2})^{2N_a}=[\frac{q_j+i(c-1)}{q_j-i(c-1)}\frac{q_j-ic}{q_j+ic}
\frac{q_j+i(c-2)}{q_j-i(c-2)}]^2\times\nonumber\\
\times\prod_{r=\pm1}\{\prod_{l\neq j}^{N-2}\frac{q_j-rq_l-i}
{q_j-rq_l+i}\prod_{\alpha=1}^M
\frac{q_j-r\lambda_\alpha+\frac i2}{q_j-r\lambda_\alpha-\frac i2}\}\\
\{\frac{\lambda_\alpha-i(c-\frac32)}{\lambda_\alpha+i(c-\frac32)}
\frac{\lambda_\alpha+i(c+\frac12)}{\lambda_\alpha-i(c+\frac12)}\}^2
\prod_{r=\pm1}\prod_{j=1}^{N-2}\frac{\lambda_\alpha-rq_j+
\frac i2}{\lambda_\alpha-rq_j-
\frac i2}=\prod_{r=\pm1}\prod_{\beta\neq\alpha}^M
\frac{\lambda_\alpha-r\lambda_\beta+i}
{\lambda_\alpha-r\lambda_\beta-i}.\nonumber
\end{eqnarray}
Following the same procedure discussed above, we obtain again a spin 
triplet ground state. This can be understood in the following picture:
 The bounded electrons and the local moments form two spin-1 local 
composites due to the effective attraction and the ferromagnetic
 Kondo coupling. However, the itinerant electrons impinging 
on these composites will screen their moments partially due
to the indirect Kondo coupling induced by the electron-electron 
correlation. In such a sense, we recover Furusaki-Nagaosa's
 conjecture\cite{8}, though the local moments are not 
completely screened.
\par
(iii) $-1/2<c<1/2$. The system falls into the antiferromagnetic Kondo
 coupling regime. No bound state appears in the ground state. By 
integrating eq.(6) in the thermodynamic limit, we have $N+2=2M$.
 This indicates a spin singlet ground state, a similar result
 to that of the Kondo problem in a Fermi liquid.

\par
(iv) $-1\leq c\leq -1/2$. No bound state exists in the ground state. 
It seems that the ground state should be a spin triplet. We note that 
both $J_a$ and $V_a$ take positive values in this region. The 
repulsive boundary potential dominates over the Kondo coupling. It
 will repel the conduction electrons to form singlet with the local 
moments and thus make the Kondo coupling ineffective. In such a sense,
 no Kondo screening occurs, which strongly indicates that both the 
boundary potential and the electron-electron correlation in the bulk 
have  significant effects to the Kondo problem in a Luttinger liquid.

\par
(v) $-3/2<c<-1$. For this case, the system is still in the regime of 
antiferromagnetic Kondo coupling but with a weaker or attractive
 boundary potential. Two imaginary $\lambda$ mode at 
$\lambda=\pm i(c+1/2)$ and two imaginary $q$ modes at 
$q=\pm i(c+1)$ appear in the ground state. These modes 
correspond to the formation of bound singlet pairs of two
 conduction electrons with the local moments. By integrating 
the densities of the real modes in the thermodynamic limit, 
we still arrive at a spin singlet ground state.

\par
(vi) $c<-3/2$. The Kondo coupling is ferromagnetic and the boundary 
potential is repulsive. There is no bound state in the ground state.
 Following the same procedure discussed in (i), we obtain again $N=2M$
 for the ground state. Therefore, there is no Kondo screening in this
 region, which contradicts to the Furusaki-Nagaosa conjecture.

\par
The thermodynamics of the present model can be calculated in a closed 
form based on the Bethe ansatz equations (4). This allows us to obtain 
the temperature and magnetic field dependence of the free energy which
 contains three parts of contributions, i.e., the bulk term, the 
boundary term and the Kondo effect term. Here we omit the detail of
 calculation following the standard procedure which can be found in 
some excellent works\cite{3,19,20,21}. The Kondo effect induced free
 energy is the most interesting one which takes the following form at
 low temperatures (Hereafter we assume $c$ is an integer or half 
integer)
\begin{eqnarray}
F_k=-Tsign(n_1)\int\frac{\ln[1+\zeta_{|n_1|}(\lambda)]d\lambda}
{2\cosh(\pi\lambda)}-Tsign(n_2)\int\frac{\ln[1+\zeta_{|n_2|}(
\lambda)]d\lambda}{2\cosh(\pi\lambda)},
\end{eqnarray}
where $n_1$ and $n_2$ are two $c$-dependent integers, 
$\zeta_{n}(\lambda)$ are elements of the following coupled 
integral equations
\begin{eqnarray}
\ln\eta(\lambda)=\frac{\epsilon_0(\lambda)-\mu}T-([1]G+[2])
\ln[1+\eta^{-1}(\lambda)]-G\ln[1+\zeta_1(\lambda)],\nonumber\\
\ln\zeta_n(\lambda)=G\{\ln[1+\zeta_{n+1}(\lambda)]+
\ln[1+\zeta_{n-1}(\lambda)],\;\;n>1,\nonumber\\
\ln\zeta_1(\lambda)=-G\ln[1+\eta^{-1}(\lambda)]+G\ln[1+
\zeta_2(\lambda)],\\
\lim_{n\to\infty}\{[n]\ln[1+\zeta_{n+1}(\lambda)]-
[n+1]\ln[1+\zeta_{n}(\lambda)]\}=\frac{2H}T,\nonumber
\end{eqnarray}
where $\mu$ is the chemical potential and 
$\epsilon_0(\lambda)=2-4\lambda^2/(4\lambda^2+1)$;
 $[n]$ and $G$ are integral operators with the kernels 
$a_n(\lambda)=(n/2\pi)/[\lambda^2+(n/2)^2]$ and
 $1/[2\cosh(\pi\lambda)]$ respectively; $H$ is the
 external magnetic field. Notice that we have omitted
 the excitations breaking the bound states in deriving 
(9) for the energy gaps associated with them. Their 
contributions to the free energy are exponentially
 small at low enough temperatures. 
\par
For $H=0$ and $T\to 0$, the driving term 
$G\ln[1+\eta^{-1}(\lambda)]$ in (9) is nothing but 
$T^{-1}\epsilon_s(\lambda)$ with $\epsilon_s(\lambda)$
 the dressed energy of the spin waves which has the 
asymptotic form $\epsilon_s(\lambda)=2\pi e_0\exp(-\pi|\lambda|)$
 for $|\lambda| \to \infty$, where $e_0$ is the energy density 
of the ground state. This allows us to formulate the 
low-temperature expansion of (9).
 The asymptotic solution of (9) is given by functions $\zeta_n(x),\;
 x=\ln[(\pi  e_0)/T]-\pi|\lambda|$ which are monotonically decreasing 
in $x$ for all n and tending to finite limits $\zeta_{n+}$ as 
$x \to \infty$. The limits are given by  
$\zeta_{n+}=\sinh^2(nH/T)/\sinh^2(H/T)-1.$
No anomaly appears in the free energy of the bulk and the open boundary. 
They contain only constant terms and $T^2$ terms up to the order $O(T^2)$.
An interesting feature is that with different $c$ value, the system may 
show both Fermi- and non-Fermi-liquid behavior. To show this, we study 
the entropy of the ground state for a variety of $c$ regions. For $c>1$ 
or $c<-3/2$, there is no bound state in the ground state. $n_1$ and $n_2$
 take the values $2|c|+1$ and $1-2|c|$ respectively. This leaves 
a residual entropy of the ground state as $S_g=\ln[(2|c|+1)/(2|c|-1)]$.
 For $c=0$, the local moments are completely screened. Both $n_1$ 
and $n_2$ are equal to unit. Thus the entropy of the ground state
 is zero and the system flows to a Fermi-liquid fixed point.
 For $c=-1/2$, $n_1=2$ and $n_2=0$. The residual entropy takes 
a value of $\ln2$. While for $c=-1$, $n_1=-1/2$, $n_2=3/2$,
 $S_g=\ln3$. Notice that for $c=-1/2$ and $c=-1$, the system 
falls into the regime of antiferromagnetic Kondo coupling  but
 with a non-zero residual entropy. We remark that $c=1/2$ and 
$c=-3/2$ are two critical points since at these points, both 
$J_a$ and $V_a$ are divergent. The residual entropy has different
 limits for $c\to1/2+0^+ (-3/2+0^+)$ and $c\to1/2+0^- (-3/2+0^-)$.
 From the above discussion we conclude that the system generally 
has a $c$-dependent residual entropy which strongly indicates a 
non-Fermi-liquid behavior. The system can flows to a Fermi-liquid 
fixed point only in a narrow parameter region of $c\sim 0$. Such
 a fascinating effect can be understood in the following picture:
 The charge-spin coupling induced by the backward scattering off 
the impurity introduces an effective boundary field to the local 
moment, which means the charge degrees of freedom join the Kondo
 effect. However, unlike a real magnetic field, the ``effective 
field" does not split the degeneracy of different orientation of
 the local moment. In fact, when the conduction electrons impinging
 and leaving the impurity, the incident waves and the reflection 
waves feel different strengths of the impurity spin. One is $1/2-c$ 
and the other is $1/2+c$ or vise versa. The finite residual entropy
 is a result of cooperative effect of the charge- and spin sectors.

\par
In summary, we introduce an integrable model of Kondo problem in a 1D 
strongly correlated electron system. With different values of 
the parameter $c$, the system can show either a Fermi liquid 
behavior or a non-Fermi liquid behavior beyond that obtained 
by Furusaki and Nagaosa\cite{8}. A non-singlet ground state 
in a antiferromagnetic Kondo coupling region is obtained. It 
is found that the residual entropy depends not only on the 
self magnetization of the ground state, but also on the 
interaction parameter $c$, which we interpret as a ``cooperative
 effect" of the Kondo coupling and the impurity scattering. It 
is instructive to apply renormalization group analysis and conformal
 field theory to reveal a fullen picture for  such an interesting 
problem. 
\par
{\bf Acknowledgment}

One of the authors (Y.W.) is supported in part by the National Natural 
Science Foundation of China and the Natural Science Foundation for Young 
Scientists, Chinese academy of Sciences.

\end{document}